\documentclass[twocolumn,english,reprint,amsmath,amssymb, onecolumn,nofootinbib,11pt]{revtex4-2}
\usepackage[T1]{fontenc}
\usepackage[latin9]{inputenc}
\usepackage{ulem}

\makeatletter
\usepackage{float}
\usepackage{physics}
\usepackage{amsfonts}
\usepackage{latexsym}
\usepackage{epsfig}
\usepackage{graphicx}
\usepackage{tikz}
\definecolor{greatblue}{RGB}{40,120,181}
\definecolor{greatred}{RGB}{200,36,35}
\usepackage[colorlinks,linkcolor=greatblue,anchorcolor=blue,citecolor=greatred]{hyperref}

\usepackage{dcolumn}% Align table columns on decimal point
\usepackage{bm}% bold math

\makeatother

\usepackage[english]{babel}

\begin{document}
\preprint{preprintnumbers}{CTP-SCU/2025011}

\title{Entanglement Entropy of Conformal Field Theory in All Dimensions}

\author{Xin Jiang}
\email{domoki@stu.scu.edu.cn}
\affiliation{College of Physics, Sichuan University, Chengdu, 610065, China}

\author{Haitang Yang}
\email{hyanga@scu.edu.cn}
\affiliation{College of Physics, Sichuan University, Chengdu, 610065, China}

\date{\today}

\begin{abstract}
We provide a field-theoretic method to calculate entanglement entropy of CFT in all dimensions.
This method works for  entangling surfaces of  arbitrary shape.  
The formalism manifests a field-theoretic
proof of the Ryu-Takayanagi formula.

%We first derive the {\it finite} entanglement entropy between disjoint subsystems in CFT$_D$. 
%The divergent entanglement entropy between adjacent subsystems is obtained by taking adjacent limits. 

%Entanglement entropy is a fundamental measure of quantum entanglement in bipartite systems. While its computation is well-established in (1+1)-dimensional conformal field theories via the replica trick, higher-dimensional cases remain challenging without resorting to holography.  In this work, we compute the vacuum entanglement entropy $S(A:B)$ between two disjoint spherical regions $A$ and $B$  in ($D\ge 3$)-dimensional conformal field theories. A key observation is that $S(A:B)$  is UV-finite and the area law for short-range entanglement emerges as a specific limit of $S(A:B)$. Our approach probably extends to arbitrarily shaped, reflection-symmetric regions.

\end{abstract}

\maketitle
\newpage
%\tableofcontents%%

\section{Introduction }

Entanglement properties of quantum systems are playing an increasingly crucial role in physics.
An intrinsic quantity counting the
degrees of freedom of entanglement is the von Neumann entropy, also
known as entanglement entropy. It is defined by $S_{A}=-\mathrm{Tr}\,\rho_{A}\log\rho_{A}$
for a reduced density matrix $\rho_{A}$ associated with a subsystem
$A$, where the total system is bipartitioned into two subsystems,
$A$ and its complement $A^{c}$. In QFT, 
one usually considers the entanglement entropy
between two {\it adjacent and complementary} subregions $A$ and $A^{c}$,
denoted as $S_\text{adj}(A:A^{c})$, where the subscript ``adj'' stands 
for adjacent entangling regions.  
Although the calculation of entanglement entropy in 
two dimensional CFT  has been well established using the replica trick 
\citep{Callan:1994py,Calabrese:2004eu}, no general approach currently exists 
for performing such calculations in higher dimensional CFT$_{D}$ for $D>2$.

\begin{figure}[h]
\includegraphics[width=0.5\textwidth]{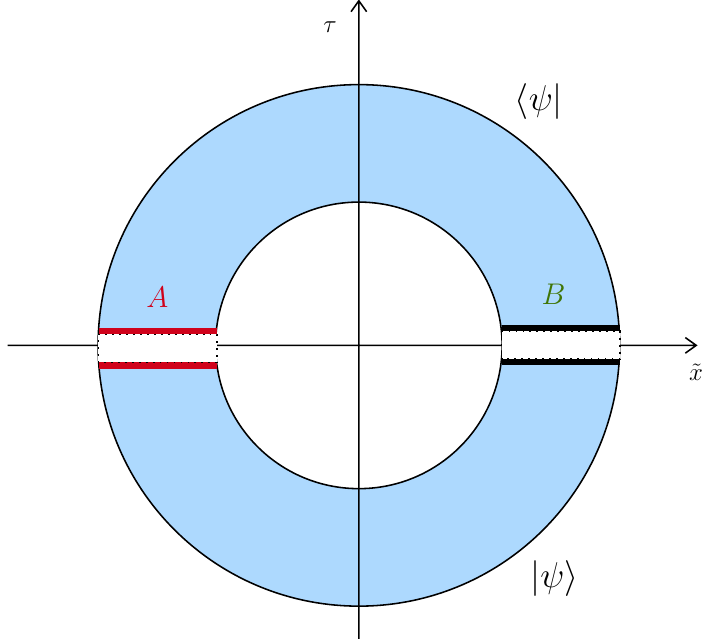}
\caption{The Euclidean pure state density matrix  $\rho =|\psi\rangle \langle\psi|$
for the annular  CFT$_2$.  
Two distinct intervals $A$ and $B$ have finite  entanglement entropy $S(A:B)$.
\label{fig:density}}
\end{figure}

\begin{figure}[h]
\includegraphics[width=0.8\textwidth]{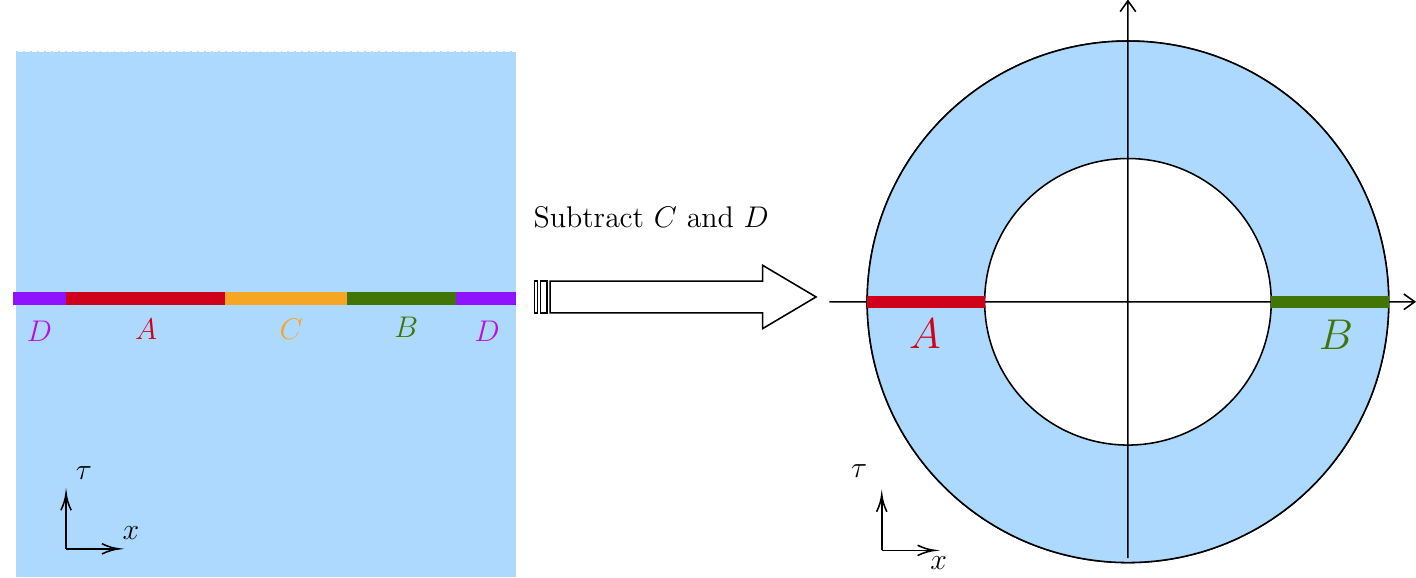}
\caption{As shown in \citep{Jiang:2024ijx,Jiang:2025tqu},  after subtracting segments $C$ and $D$
with two discs in the infinite system, we obtain an  annular region in
which $A$ and $B$ are in a pure entangled state $\psi_{AB}$. 
It is identical to the annular CFT$_2$ in Figure \ref{fig:density}.
\label{fig:subtraction}}
\end{figure}

The entropy $S_\text{adj}(A:A^{c})$ typically suffers  an
ultraviolet (UV) divergence due to the very intense entanglement between
contiguous fields. This  divergent behavior renders it extremely challenging, 
if not entirely impossible,
to derive exact relations between QFT and gravity. 
However, for a UV-complete theory such as a CFT, 
all physical quantities must possess well-defined UV-finite forms. 
Divergences and regulators, 
rather than being intrinsic components of the theory itself, 
should only emerge as artifacts of specific limiting procedures.
A natural question thus arises: Do there exist any finite {\it elementary} 
entanglement entropies in CFT? Here, the term {\it elementary} is used in contrast to 
{\it compound} quantities such as mutual information, relative entropy, 
and the like --- which are constructed from combinations of elementary entanglement measures.
For such finite elementary entanglement entropies  to exist,  two conditions should be satisfied:
\begin{itemize}
\item The entangled regions must be disjoint.
\item The system  has to be a pure state to allow the calculation of the  entanglement entropy.
\end{itemize}
The annular CFT$_2$ in Figure \ref{fig:density} precisely meets these two conditions.
Note the time direction is upward. The lower half annulus corresponds to $|\psi\rangle$,
and the upper half  annulus corresponds to $\langle\psi|$. The density matrix of this pure state is 
 $\rho =|\psi\rangle \langle\psi|$. 
In prior collaborations \citep{Jiang:2024ijx,Jiang:2025tqu,Jiang:2025dir}, 
we demonstrated  that the entanglement entropy $S_\text{disj}(A:B)$  
between disjoint intervals $A$ and $B$ in this annular CFT is indeed finite, for static, covariant and 
thermal configurations, respectively. The subscript ``disj'' denotes  disjoint entangling regions.
Notably,  referring to  Figure \ref{fig:subtraction}, as shown in \citep{Jiang:2024ijx,Jiang:2025tqu},
$S_\text{disj}(A:B)$ can be equivalently interpreted as the entanglement entropy 
between two disjoint intervals  in the infinite system.
The well-known divergent entanglement entropies $S_\text{adj}(A:B)$ for adjacent intervals emerge 
naturally as the adjacent limits of disjoint $S_\text{disj}(A:B)$.

A critical insight  is the straightforward generalization of the annular CFT$_2$ 
to higher dimensions:   CFT$_D$  on a solid torus  $\mathbb{B}^{D-1}\times S^1$, 
see  Figure \ref{fig:solid-torus}, where $\mathbb{B}^{D-1}$ is a $(D-1)$-dimensional ball. 
Since the replica of a  solid torus is still a  solid torus, 
calculating entanglement entropies simplifies to computing the partition function on this solid torus!
The usual divergent entanglement entropies for adjacent regions 
are again easily obtained by taking adjacent limits. 
In this paper, we only study the ball shaped entangling region. But one can easily see 
that this framework  works for entangling surfaces of arbitrary shape.

For $D>2$, the relationship between the solid torus CFT$_D$ and the infinite 
CFT$_D$ on $\mathbb{R}^D$ mirrors the $D=2$ case. Consider calculating the 
entanglement entropy between two codimension-one disjoint discs $A$ and $B$ in 
\(\mathbb{R}^D\). 
%corresponding to the mixed state \(\rho_{AB}\) of the infinite system. 
Analogous to the two-dimensional scenario, removing a $D$-dimensional solid torus 
between $A$ and $B$ yields also a $D$-dimensional solid torus depicted in Figure 
\ref{fig:solid-torus} --- precisely the SUBTRACTION procedure introduced in \citep{Jiang:2024ijx}.

%
%
%Imagine we want to 
%calculate the entanglement entropy between two co-dimension one 
%disjoint discs $A$ and $B$ on  $R^D$, which is usually interpreted as 
%the mixed state $\rho_{AB}$ of  $R^D$. Just as  $D=2$ case, we remove a D-dimensional 
%solid torus between $A$ and $B$, obtaining the  solid torus  in Figure \ref{fig:solid-torus}.
%This is precisely the  SUBTRACTION approach introduced in \citep{Jiang:2024ijx}. 

The remainder of this paper is outlined as follows.
In section II, we derive the explicit expression of the entanglement entropy between 
two disjoint balls in all dimensions.
In section III, by taking the adjacent limits,  the  divergent entanglement 
entropies for adjacent balls in all dimensions are obtained,
where the area law \citep{Sorkin1983,Srednicki1993} is manifested. 
In section IV, we provide examples in CFT$_2$ and CFT$_4$.
In section V, we show how RT formula \citep{Ryu:2006bv,Ryu:2006ef} 
naturally emerges in this framework and discuss arbitrary shape of entangling surfaces.
Section VI is the conclusion.

\section{Entanglement entropy between two disjoint balls}

%For a CFT in $D$-dimensional Euclidean spacetime with the metric
%$\mathrm{d}s_{0}^{2}=\mathrm{d}t_{\text{E}}^{2}+\mathrm{d}y^{2}+  \underset{i=1}{\overset{D-2}{\sum}}
%\mathrm{d}x_{i}^{2}$,

\begin{figure}[h]
\includegraphics[scale=0.6]{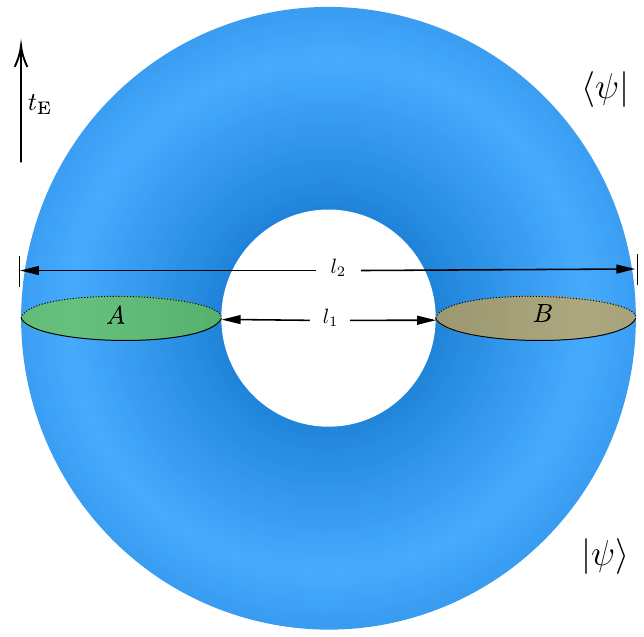}
\caption{The Euclidean pure state density matrix $\rho = |\psi\rangle \langle\psi|$ for the solid torus CFT$_D$.
The lower half solid torus corresponds to $|\psi\rangle$.
The upper half  annulus corresponds to $\langle\psi|$. 
We calculate the entanglement entropy between two disjoint co-dimensional one balls $A$ and $B$ (colored regions) in the bounded
region $\mathcal{B}$ (blue region).
\label{fig:solid-torus}}
\end{figure}

Let us consider the CFT$_D$  on a solid torus, 

\begin{equation}
\mathcal{B} =\mathbb{B}^{D-1}\times S^{1} :=\left\{ (t_{\text{E}},y,x)\vert\left(\sqrt{t_{\text{E}}^{2}+y^{2}}-\frac{l_{2}+l_{1}}{2}\right)^{2}+
\sum_{i=1}^{D-2} x_{i}^{2}<\left(\frac{l_{2}-l_{1}}{2}\right)^{2}\right\},
\end{equation}
where $\mathbb{B}^{D-1}$ is a $(D-1)$-dimensional ball and $l_{2}>l_{1}>0$. 
At the time slice $t_{\text{E}}=0$, $\mathcal{B}$
consists of two disjoint $(D-1)$-balls $A$ and $B$ with the same
radius, as shown in Figure \ref{fig:solid-torus}. 
The lower half solid torus corresponds to $|\psi\rangle$,
and the upper half  annulus corresponds to $\langle\psi|$. 
The density matrix of this pure state is  $\rho =|\psi\rangle \langle\psi|$. 

\begin{figure}[h]
\begin{centering}
\includegraphics[width=0.5\textwidth]{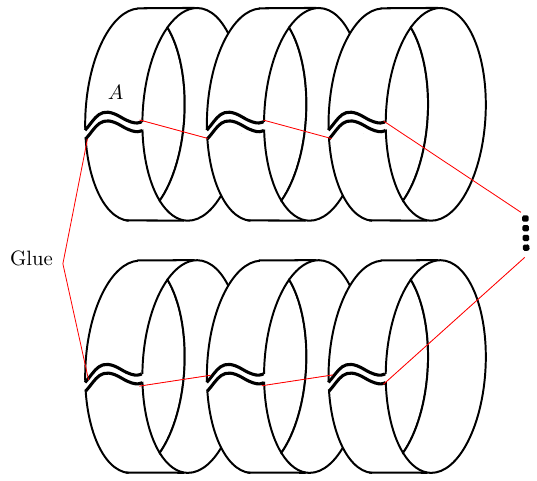}
\par\end{centering}
\caption{\label{fig:replica}The cut-and-glue procedure
in the replica trick to construct  $\mathcal{B}_{n}$. 
Each solid torus is cut along the $(D-1)$-dimensional ball $A$
and is glued with others cyclically. Red lines represent gluing operations.
Note that the resulted manifold is also a solid torus with period $2n\pi$.}
\end{figure}

%The bounded region
%$\mathcal{B}$ is intuitively a solid torus $\mathbb{B}^{D-1}\times S^{1}$,
%where $\mathbb{B}^{D-1}$ is a $(D-1)$-dimensional ball.
Using the replica trick \citep{Callan:1994py}, 
the entanglement entropy $S_\text{disj}(A:B)$  between $A$ and $B$ can be defined 
through  the R\'{e}nyi entropy as

%Starting point of our computations is the observation that the 
%entanglement entropy between $A$ and $B$ can be computed by the
%replica trick \citep{Callan:1994py}. Such techniques allow us to
%write the R\'{e}nyi entropy as
\begin{eqnarray}
&&S^{(n)} (A:B) = \frac{1}{1-n} \log \mathrm{Tr}_A \rho_A^n =\frac{1}{1-n}\log\left[\frac{Z_{\mathcal{B}_{n}}}{Z^{n}}\right],\nonumber\\
&&S_\text{disj}(A:B) = \lim_{n\to 1}S^{(n)} (A:B),
\label{eq:Renyi}
\end{eqnarray}
where $\rho_A =\mathrm{Tr}_B\rho$ is the reduced density matrix.  
$Z$ is the partition function on the solid torus  $\mathcal{B}$.  $Z_{\mathcal{B}_{n}}$ is the 
partition function on the $n-$sheeted cover  $\mathcal{B}_{n}$ made by cyclically gluing $n$ copies
of $\mathcal{B}$  along $A$ (or equivalently $B$), see Figure \ref{fig:replica} where the cut $A$ is a
$(D-1)$-dimensional ball $\mathbb{B}^{D-1}$.

Since  $\mathcal{B}_{n}$ is still a solid torus with period 
$2n\pi$, calculating the $S_\text{disj}(A:B)$ in equation (\ref{eq:Renyi}) is simplified to  
computing the partition function on a solid torus $\mathcal{B} =\mathbb{B}^{D-1}\times S^{1}$!
A brute force approach to this calculation would involve solving the equations of motion (EoM) 
with either Dirichlet or Neumann boundary conditions
which makes no difference since the vacuum entanglement is concerned. 
Instead, using the fact that a periodic system is equivalent to a thermal system,
by noting the solid torus  $\mathcal{B}$ is periodic,
we adopt a symmetry based method: this approach is universal across all
CFTs, applicable to  free or strongly coupled systems. 

%Since the solid torus  $\mathcal{B}$ is periodic, the CFT on it is equivalent to a thermal system. 
Applying the coordinate transformation
$t_{\text{E}}=r\sin\theta$, $y=r\cos\theta$ to the flat metric $\mathrm{d}s_{0}^{2}=\mathrm{d}t_{\text{E}}^{2}+\mathrm{d}y^{2}+  \underset{i=1}{\overset{D-2}{\sum}}
\mathrm{d}x_{i}^{2}$,  we get a conformally flat metric
\[ \mathrm{d}s_0^{2}=r^2\left[\mathrm{d}\theta^{2}+\Big(\mathrm{d}r^{2}+\sum_{i=1}^{D-2} 
\mathrm{d}x_{i}^{2}\Big)/r^{2} \right].
\]
Removing the conformal factor $r^2$ by the Weyl invariance, the spacetime metric becomes
\begin{equation}
\mathrm{d}s^{2}=\mathrm{d}\theta^{2}+\left(\mathrm{d}r^{2}+\sum_{i=1}^{D-2} 
\mathrm{d}x_{i}^{2}\right)/r^{2}.\label{eq:rad-coord}
\end{equation}
Since $\theta$ plays a role of the imaginary time with a periodicity $2\pi$,
this CFT is equivalent to a thermal system with the inverse temperature $\beta =2\pi$.
It is straightforward to write down
the partition function of this thermal system
\begin{equation}
Z=\mathrm{Tr}\,e^{-2\pi H},
\end{equation}
where the Hamiltonian operator $H$ is conjugate to the imaginary time $\theta$, 
defined by
\[
H=\int_{\mathbb{B}^{D-1}}T^{\theta\theta}
\]
where $T$ is the stress-energy tensor. By noting the periodicity of the $n-$sheeted cover  $\mathcal{B}_{n}$ 
is $2n\pi$, we immediately get
the partition function on $\mathcal{B}_{n}$,
\begin{equation}
Z_{\mathcal{B}_{n}}=\mathrm{Tr}\,e^{-2n\pi H_{(n)}},
\end{equation}
with the corresponding Hamiltonian operator $H_{(n)}$. We will focus
on the partition function that is dominated by the ground state, 
\begin{equation}
Z\simeq e^{-2\pi\langle H\rangle},\quad Z_{\mathcal{B}_{n}}\simeq e^{-2n\pi\langle H_{(n)}\rangle}.
\end{equation}
$\langle H\rangle$ denotes the vacuum expectation value of $H$.
Note that $\mathcal{B}$ is related to the replicated manifold $\mathcal{B}_{n}$
by a rescaling of imaginary time $\theta\rightarrow n\theta$, which
implies that\footnote{Note the stress-energy tensor is  covariant
under  coordinates transformations, $T^{\theta'\theta'}\partial_{\theta'}\partial_{\theta'}=T^{\theta\theta}\partial_{\theta}\partial_{\theta}$,
and $\theta'=n\theta$.}
\begin{equation}
\langle H_{(n)}\rangle=\frac{\langle H\rangle}{n^{2}}.
\end{equation}
Then, the R\'{e}nyi entropy in equation (\ref{eq:Renyi}) becomes
\begin{equation}
S^{(n)}(A:B)=\frac{1}{1-n}\log\left[\frac{e^{-2\pi\langle H\rangle/n}}{e^{-2n\pi\langle H\rangle}}\right]=-2\pi\left(1+\frac{1}{n}\right)\langle H\rangle.
\end{equation}
The entropy $S_\text{disj}(A:B)$ of entanglement between two disjoint balls $A,B$
therefore reads
\begin{equation}
S_\text{disj}(A:B)=S^{(1)}(A:B)=-4\pi\langle H\rangle.
\label{eq:S1}
\end{equation}
We stress that the Hamiltonian $H$ is conjugate to $\theta$ but
not to $t_{\text{E}}$, so it is actually dimensionless.

The entanglement entropy $S_\text{disj}(A:B)$ simply depends on the ground state
energy $\langle H\rangle$ localized in a spatial ball $\mathbb{B}^{D-1}$:
\begin{equation}
\langle H\rangle=\int_{\mathbb{B}^{D-1}}\mathcal{E}_{\text{vac}},
\label{eq:Hintegral}
\end{equation}
where $\mathcal{E}_{\text{vac}}$ is the Casimir energy density that
is usually a negative constant in our discussions. The integration is performed
with the metric of equation (\ref{eq:rad-coord}).
Consequently, the
entanglement entropy $S_\text{disj}(A:B)$ is given by
\begin{eqnarray}
S_\text{disj}(A:B) & = & -4\pi\mathcal{E}_{\text{vac}}\mathrm{Vol}(\mathbb{B}^{D-1}),\nonumber\\
 & = & -4\mathcal{E}_{\text{vac}}\pi^{\frac{D}{2}}\frac{\Gamma\left(\frac{D}{2}\right)}{\Gamma\left(D\right)}\left(\frac{l_{2}}{l_{1}}-1\right)^{D-1}\,_{2}F_{1}\left(D-1,\frac{D}{2};D;1-\frac{l_{2}}{l_{1}}\right).
 \label{eq:gen-result}
\end{eqnarray}
with the standard hypergeometric function $\,_{2}F_{1}$. The parameters $l_1$ and $l_2$ are 
the inner and outer radii of the solid torus.

\begin{figure}[h]
\begin{centering}
\includegraphics{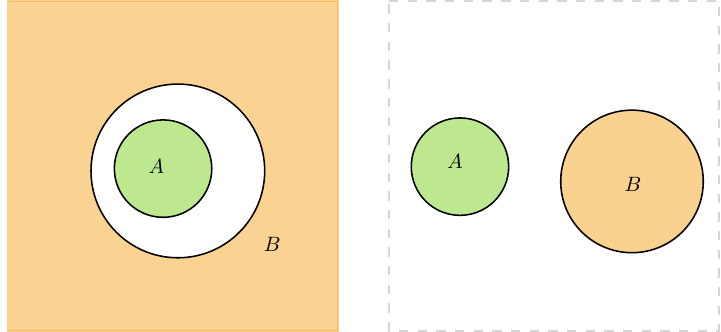}
\par\end{centering}
\caption{Entangling regions (colored regions) are bounded by ($D-2$)-spheres
(black circles), respectively. Left panel (the cavity configuration):
A ($D-1$)-ball living in a spheric cavity is entangled with the region
outside the cavity. Right panel (the juxtaposed configuration): Two disjoint
($D-1$)-balls with different radii are entangled.\label{fig:Entangling-regions}}
\end{figure}

Generally, there exist two different configurations of our interest,
namely, the cavity configuration and the juxtaposed configuration illustrated
in Figure \ref{fig:Entangling-regions}. Each configuration
is uniquely fixed by boundaries of $A,B$, i.e., two disjoint
($D-2$)-spheres. Every ($D-1$)-ball is bounded by a ($D-2$)-sphere
$\Sigma(\vec{x},r)$ located at $\vec{x}\in\mathbb{R}^{D-1}$ with
the radius $r$, respectively. An elegant substitute for conformality
is the inversive product $\varrho$ of two spheres $\Sigma(\vec{x},r)$
and $\Sigma(\vec{x}^{\prime},r^{\prime})$ \citep{beardon2012geometry}
\begin{equation}
\varrho=\left|\frac{r^{2}+r^{\prime2}-\vert\vec{x}-\vec{x}^{\prime}\vert^{2}}{2rr^{\prime}}\right|.
\label{eq:inversive product}
\end{equation}
In the case that two disjoint ($D-1$)-balls $A,B$ share the same
radius, we have
\begin{equation}
\frac{l_{2}}{l_{1}}=\frac{\sqrt{\varrho+1}+\sqrt{2}}{\sqrt{\varrho+1}-\sqrt{2}}.\label{eq:univ-ratio}
\end{equation}

The inversive product $\varrho$ is a {\it conformal invariant},
as it remains invariant under global conformal transformations. 
If two configurations have different inversive products, they cannot
be conformally mapped onto each other. The set of all configurations
falls into classes of conformally equivalent regions, with each class
characterized by the inversive product $\varrho$ of that class. 
Consequently, any two configurations sharing the same $\varrho$ 
must be conformally equivalent.
To compute the entanglement entropy $S_\text{disj}(A:B)$ for any two disjoint spheres in general 
configurations, it suffices to evaluate the inversive product $\varrho$ of the spheres. 
Thus, we conclude that the entanglement entropy between any two 
disjoint spheres is given 
by
\begin{equation}
S_\text{disj}(A:B)=-4\mathcal{E}_{\text{vac}}\pi^{\frac{D}{2}}\frac{\Gamma\left(\frac{D}{2}\right)}{\Gamma\left(D\right)}\left(\frac{2\sqrt{2}}{\sqrt{\varrho+1}-\sqrt{2}}\right)^{D-1}\,_{2}F_{1}\left(D-1,\frac{D}{2};D;\frac{2\sqrt{2}}{\sqrt{2}-\sqrt{\varrho+1}}\right).\label{eq:Gen-ee}
\end{equation}
The time dependent covariant scenario is obtained by simply replacing $\vec x$ with $x^\mu$.

\section{Emergent area law in various dimensions\label{sec:Emergent-area-law}}

In this section, we take the adjacent limit of the general finite entanglement entropy (\ref{eq:Gen-ee})
of disjoint balls to 
give the usual divergent  entanglement entropies of adjacent balls.
\begin{figure}[h]
\includegraphics{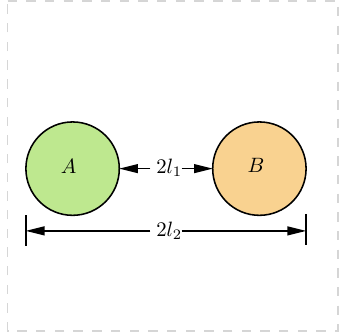}$\qquad$\includegraphics{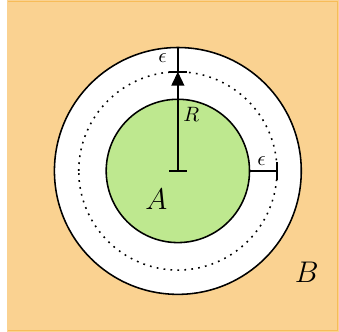}
\caption{Left panel: The time zero slice of $\mathcal{B}$ in $D$ dimensions,
where $A$ and $B$ are two disjoint ($D-1$)-balls with the same
radius (colored regions) living at $t_{\text{E}}=0$. Right panel:
The ($D-1$)-ball $A$ with the radius $R-\epsilon$ is in a spheric
cavity with the radius $R+2\epsilon$, and the region $B$ is outside
the cavity. Here, $\epsilon$ plays a role of the UV cutoff.\label{fig:A-and-B}}
\end{figure}
We first consider the entanglement entropy $S_\mathrm{disj}(A:B)$ for two disjoint
($D-1$)-balls $A,B$ with the same radius  in the juxtaposed configuration, as shown in Figure \ref{fig:A-and-B}.
It would be convenient to write down $S_\text{disj}(A:B)$ in various dimensions:
\begin{eqnarray}
D=2: &\quad  & 4\pi\mathcal{E}_{2}\log\frac{l_{2}}{l_{1}},\nonumber \\
D=3: &  & 4\pi^{2}\mathcal{E}_{3}\left(\sqrt{\frac{l_{2}}{l_{1}}}+\sqrt{\frac{l_{1}}{l_{2}}}\right)-8\pi^{2}\mathcal{E}_{3},\nonumber \\
D=4: &  & 2\pi^{2}\mathcal{E}_{4}\left(\frac{l_{2}}{l_{1}}-\frac{l_{1}}{l_{2}}\right)-4\pi^{2}\mathcal{E}_{4}\log\frac{l_{2}}{l_{1}},\nonumber \\
D=5: &  & \frac{\pi^{3}\mathcal{E}_{5}}{3}\left[\left(\frac{l_{2}}{l_{1}}\right)^{3/2}+\left(\frac{l_{1}}{l_{2}}\right)^{3/2}-9\left(\sqrt{\frac{l_{2}}{l_{1}}}+\sqrt{\frac{l_{1}}{l_{2}}}\right)\right]+\frac{16\pi^{3}\mathcal{E}_{5}}{3},\nonumber \\
D=6: &  & \frac{\pi^{3}\mathcal{E}_{6}}{6}\left[\left(\frac{l_{2}}{l_{1}}\right)^{2}-\left(\frac{l_{1}}{l_{2}}\right)^{2}-8\left(\frac{l_{2}}{l_{1}}-\frac{l_{1}}{l_{2}}\right)\right]+2\pi^{3}\mathcal{E}_{6}\log\frac{l_{2}}{l_{1}},\label{eq:D-EE}
\end{eqnarray}
where we denote $\mathcal{E}_{\text{vac}}=-\mathcal{E}_{D}<0$ and
remember that $\mathcal{E}_{D}$ is actually different in various
dimensions. 

Then, we return to the general $D$-dimensional cavity configuration
depicted in the right panel of Figure \ref{fig:A-and-B}, where $A$
is almost adjacent to $B$, more precisely
\begin{equation}
\vec{x}=\vec{x}^{\prime},\quad r=R+\epsilon,\quad r^{\prime}=R-\epsilon.
\end{equation}
In this setup, the inversive product becomes
\begin{equation}
\varrho=\frac{R^{2}+\epsilon^{2}}{R^{2}-\epsilon^{2}}.
\end{equation}
Plugging $\varrho$ into the general expression (\ref{eq:Gen-ee}), we obtain  $S_\text{adj}(A:B)$ in various dimensions:
\begin{eqnarray}
D=2: & \quad & 8\pi\mathcal{E}_{2}\log\frac{2R}{\epsilon}+\mathcal{O}(\epsilon),\nonumber \\
D=3: &  & 8\pi^{2}\mathcal{E}_{3}\left(\frac{R}{\epsilon}-1\right),\nonumber \\
D=4: &  & 2\pi^{2}\mathcal{E}_{4}\left(\frac{4R^{2}}{\epsilon^{2}}-2-4\log\frac{2R}{\epsilon}\right)+\mathcal{O}\left(\epsilon\right),\nonumber \\
D=5: &  & \frac{8\pi^{3}\mathcal{E}_{5}}{3}\left(\frac{R^{3}}{\epsilon^{3}}-\frac{3R}{\epsilon}+2\right),\nonumber \\
D=6: &  & \frac{\pi^{3}\mathcal{E}_{6}}{3}\left(\frac{8R^{4}}{\epsilon^{4}}-\frac{24R^{2}}{\epsilon^{2}}+9+12\log\frac{2R}{\epsilon}\right)+\mathcal{O}\left(\epsilon\right),\label{eq:D-EE-UV}
\end{eqnarray}
which perfectly agrees with holographic predictions given in  \citep{Ryu:2006bv}.
In $D=2$, the result is the well-known $\log$-law for entanglement
entropy. In $D\ge3$, we can immediately observe the area law of entanglement
entropy, which arises as a special limit of our general consideration.
We emphasize  that in even dimensions, the finite term
is physically irrelevant since they can be absorbed into the UV cutoff $\epsilon$.
{\it But in odd dimensions, finite terms of entanglement entropies become scheme-independent and
play a crucial role in the F-theorem} \citep{Jafferis:2010un,Jafferis:2011zi,Liu2013}. {\it
This is manifested by the exactness
of our expressions for entanglement entropy in odd  dimensions.}

\section{Examples in CFT$_{2}$ and CFT$_{4}$}

We provide two  checks in this section.
Consider a CFT$_{4}$ living in the 4-dimensional conformally flat
spacetime with the metric $\mathrm{d}s^{2}=\mathrm{d}\theta^{2}+\left(\mathrm{d}r^{2}+\sum_{i=1}^2 \mathrm{d}x_{i}^{2}\right)/r^{2}$,
where the vacuum expectation value of the stress tensor has been explicitly
calculated in \citep{Brown:1977sj}, 
%The Casimir energy density can be directly obtained,
\begin{equation}
\mathcal{E}_{\text{vac}}=\left\langle T^{\theta\theta}\right\rangle =-\frac{3a+c}{8\pi^{2}},\label{eq:4d-energy}
\end{equation}
where $a$ and $c$ are different coefficients of the trace 
anomaly\footnote{In our notations, the trace anomaly is $\left\langle T_{\;\mu}^{\mu}\right\rangle =\frac{1}{(4\pi)^{2}}\left[c(C_{\mu\nu\rho\sigma}^{2}+\frac{2}{3}\nabla^{2}\mathcal{R})-a\mathcal{G}\right]$,
with the Weyl tensor $C_{\mu\nu\rho\sigma}$, the Ricci scalar $\mathcal{R}$
and the topological Euler density $\mathcal{G}.$}. 
In the 4-dimensional $\mathcal{N}=4$ $SU(N)$ super Yang-Mills gauge
theory, the anomaly coefficients are \citep{Birrell:1982ix,Gubser:1997se}
\begin{equation}
a=\frac{N^{2}}{4},\quad c=\frac{N^{2}}{4}.
\end{equation}
%\sout{\textcolor{blue}{Plugging Equation (\ref{eq:4d-energy}) and $D=4$ into Equation
%(\ref{eq:gen-result}), the entanglement entropy between $A$ and
%$B$ is given by
%\begin{equation}
%S_\text{disj}(A:B)=\frac{3a+c}{4}\left(\frac{l_{2}}{l_{1}}-\frac{l_{1}}{l_{2}}\right)-\frac{3a+c}{2}\log\frac{l_{2}}{l_{1}}.
%\end{equation}
%Note that $A$, $B$ are two disjoint 3-balls with the same radius.
%In the 4-dimensional $\mathcal{N}=4$ $SU(N)$ super Yang-Mills gauge
%theory, the anomaly coefficients are \citep{Birrell:1982ix,Gubser:1997se}
%\begin{equation}
%a=\frac{N^{2}}{4},\quad c=\frac{N^{2}}{4}.
%\end{equation}
%}}
Hence from equation (\ref{eq:D-EE}), for the juxtaposed configuration, 
\begin{equation}
S_\text{disj}(A:B)=\frac{N^{2}}{4}\left(\frac{l_{2}}{l_{1}}-\frac{l_{1}}{l_{2}}\right)-\frac{N^{2}}{2}\log\frac{l_{2}}{l_{1}}.
\end{equation}

\noindent For the right panel of Figure \ref{fig:A-and-B},
from equation (\ref{eq:D-EE-UV}), the adjacent entanglement entropy 
$S_\text{adj}(A:A^c)$ of a $3-$ball in  $\mathbb{R}^3$ is

\begin{equation}
S_\text{adj}(A:A^{c})\simeq N^{2}\left[\left(\frac{R}{\epsilon}\right)^{2}-\log\frac{2R}{\epsilon}-2\right]+\mathcal{O}(\epsilon),
\end{equation}
which precisely matches the holographic prediction given in \citep{Ryu:2006bv}. 

As $D=2$, the metric reduces to $\mathrm{d}s^{2}=\mathrm{d}\theta^{2}+\mathrm{d}r^{2}/r^{2}$, 
which is just a flat metric on the Euclidean cylinder. 
Now, $A=\{(\theta,r)\vert l_{1}<r<l_{2},\theta=\pi\}$
and $B=\{(\theta,r)\vert l_{1}<r<l_{2},\theta=0\}$ are two disjoint
intervals. The Casimir energy density is related to the central
charge \citep{DiFrancesco:1997nk}: 
\begin{equation}
\mathcal{E}_{\text{vac}}=-\frac{c}{24\pi}.
\end{equation}
The CFT$_{2}$ result presented in our previous works \citep{Jiang:2024ijx,Jiang:2025tqu} is immediately reproduced,
\begin{equation}
S(A:B)=\frac{c}{6}\log\frac{l_{2}}{l_{1}}.
\end{equation}
Its adjacent limit gives the entanglement entropy of a single interval
with the length $2R$,
\[
S_\text{adj}(A:A^{c})=\frac{c}{3}\log\frac{2R}{\epsilon}+\mathcal{O}(\epsilon).
\]
%which is consistent with the seminal work \citep{Callan:1994py,Calabrese:2004eu}.

\section{Ryu-Takayanagi formula and Shape independence}
We present two important indications or extensions in this section.

\subsection{Manifestation of Ryu-Takayanagi formula}

Our derivation has already manifested the RT formula. 
From the first line of equations (\ref{eq:gen-result}), 
for a symmetric solid torus (reflection-symmetric juxtaposed configuration) 
as illustrated by  the left panel of Figure \ref{fig:A-and-B},
the entanglement entropy between balls $A$ and $B$ is the volume  Vol$(A)$ 
(or equivalently Vol$(B)$) calculated in the hyperbolic metric 
$\mathrm{d}s^{2}=(\mathrm{d}r^{2}+\underset{i=1}{\overset{D-2}{\sum}}\mathrm{d}x_{i}^{2})/r^{2}.$
This hyperbolic volume  Vol$(A)$  could always be understood as a
($D-1$)-dimensional minimal surface embedded in AdS$_{D+1}$ space, 
which is precisely the entanglement wedge cross-section (EWCS), i.e RT surface,
for this symmetric configuration.

As for a generic disjoint configurations  in Figure \ref{fig:Entangling-regions},
we simply make a conformal transformation to get the corresponding RT surface (EWCS). 
This actually provides a simple way to compute the minimal surface for a particular
configuration. It is of interest to verify the result in a geometric picture.

When taking the adjacent limit, one immediately reproduces the original RT formula for  
the  adjacent scenario with divergent entanglement entropy.

It is very interesting to note that, 
the disjoint entanglement entropy  $S_\text{disj}(A:B)$ is proportional to the {\it hyperbolic} volume of the
entangling region., rather than the area of the entangling surface.
As taking the adjacent limit, the contribution from the entangling surface dominates and the area law emerges.

\subsection{Shape independence}

Although we only study the spherical entangling surface in this paper, 
our approach is clearly independent of the specific  shape of entangling regions $A$ and $B$
in the reflection-symmetric juxtaposed configuration. 
One simply replaces $A$ and $B$ with a generally shaped region,
denoted as $\mathbb{X}$. The  entanglement entropy would
then be
\begin{eqnarray}
S_\text{disj}(A:B) & = & -8\pi\mathcal{E}_{\text{vac}}\mathrm{Vol}(\mathbb{X}),\label{eq:gen-result-1}
\end{eqnarray}
where $\mathrm{Vol}(\mathbb{X})$ is easily calculated with the conformal metric (\ref{eq:rad-coord}).

%A particular interesting object is the divergent adjacent entanglement entropy  
%$S_\text{adj}(\mathbb{Y}:\mathbb{Y}^c)$ for an arbitrarily shaped region  $\mathbb{Y}$
%in $\mathbb{R}^D$.
%This object is obtained as the adjacent limit of $S_\text{disj}(\mathbb{Y}:\mathbb{Y}^c)$ 
%in the cavity configuration, where  inner  and outer
%regions share the same desired shape $\mathbb{Y}$ but differ in size.  
%To calculate $S_\text{disj}(\mathbb{Y}:\mathbb{Y}^c)$, we need to map it
%to the juxtaposed configuration, which nevertheless typically ends up with a
%non-reflection-symmetric juxtaposed configuration.
%Since the thermal partition function usually cannot be formulated for 
%generic non-reflection-symmetric juxtaposed configurations, 
%analytic expression of 
%$S_\text{disj}(\mathbb{Y}:\mathbb{Y}^c)$ may not be available and we have
%to turn to numerical computation.

A particular interesting object is the divergent adjacent entanglement entropy  
$S_\text{adj}(\mathbb{Y}:\mathbb{Y}^c)$ for an arbitrarily shaped region  $\mathbb{Y}$
in $\mathbb{R}^D$.
It can be calculated as  follows:

\begin{itemize}
\item We start with $S_\text{disj}(\mathbb{Y}_1:\mathbb{Y}_2)$
in the cavity configuration. $\mathbb{Y}_1$ and $\mathbb{Y}_2$ are
nested,  centered at the same point, different in scale.

\item To calculate $S_\text{disj}(\mathbb{Y}_1:\mathbb{Y}_2)$, we need to map it
to the juxtaposed configuration.

\item  Typically we end up with a
non-reflection-symmetric juxtaposed configuration.

\item Since the thermal partition function usually cannot be formulated for 
generic non-reflection-symmetric juxtaposed configurations, 
analytic expression of $S_\text{disj}(\mathbb{Y}_1:\mathbb{Y}_2)$ may not be available. 
In this case, we can   compute  the partition function on the 
non-reflection-symmetric torus numerically.

\item Finally,  take the adjacent limit,
\[S_\text{adj}(\mathbb{Y}:\mathbb{Y}^c) = \lim_{\mathbb{Y}_1 \to \mathbb{Y}_2} S_\text{disj}(\mathbb{Y}_1:\mathbb{Y}_2).\]

\end{itemize}

\section{Conclusion}

In summary, we introduced a general  framework to compute the  entanglement entropy in CFT$_D$ for $D\geq 2$. 
We provided an explicit expression for the  entanglement entropy of ball-shaped entangling regions in CFT$_D$.
This method is applicable to entangling regions of arbitrary shape. 
Remarkably, this method is substantially   simpler
than the traditional one used to calculate CFT$_2$ entanglement entropy. 
The RT formula emerges naturally within this formalism.

In \citep{Hertzberg:2010uv}, some parts of finite contributions to the adjacent entanglement entropy
were extracted for $D > 2$ under specific limits, including: 
(i) entropy between the interior and exterior of a spatial domain for a massive scalar field;
(ii) numerically computed entropy for an interval in CFT.  
Extending our framework to massive QFT  is a critical research avenue.
Additionally, \citep{Casini:2011kv} established the universal logarithmic term 
of adjacent entanglement entropy in even-dimensional CFT. 
A particularly intriguing direction is to recover the hyperbolic mapping 
introduced in that work from the adjacent limits of our results.

In CFT\(_2\), the divergent entanglement entropy between two adjacent and complementary 
subsystems has been extensively studied. However, the finite entanglement entropy 
between two disjoint regions has remained underexplored until recently.
From our perspective, the latter is of greater fundamental importance: 
it provides a pair of exactly equal dual quantities in the correspondence 
between gravity and CFT, whereas the divergent adjacent configuration represents 
merely a special limiting case of this disjoint configuration.

\vspace*{3.0ex}
\begin{acknowledgments}
\paragraph*{Acknowledgments.} 
This work is supported by NSFC (Grant No. 12275184).
\end{acknowledgments}

\bibliographystyle{unsrturl}
\bibliography{ref202510}

\end{document}